\documentclass[12pt]{article}

\advance\voffset by -0.25cm

\usepackage{graphicx,amsmath,amsfonts,amsthm,bbm,setspace,url}
\usepackage{natbib,verbatim,mathrsfs}

\usepackage{mymacrosbeamer}

\begin{document}
\baselineskip=20pt

\begin{center}{
  \Large \bf
  Stochastic Tropical Cyclone Precipitation Field Generation
}

\bigskip
\bigskip

{\bf William Kleiber\footnote[1]{Department of Applied Mathematics,
  University of Colorado, Boulder, CO.  
  Corresponding author e-mail: \texttt{william.kleiber@colorado.edu}},
  Stephan Sain\footnote[2]{Jupiter Intelligence, Boulder, CO.},
  Luke Madaus\footnotemark[2]
  and 
  Patrick Harr\footnote[3]{Jupiter Intelligence, San Mateo, CA.}
}

\bigskip
\bigskip

{\bf \today}

\end{center}

\bigskip
\bigskip

\begin{abstract}
Tropical cyclones are important drivers of coastal flooding which have severe negative public safety and economic consequences. 
Due to the rare occurrence of such events, high spatial and temporal resolution historical storm precipitation data are limited in availability. 
This paper introduces a statistical tropical cyclone space-time precipitation generator given limited information from storm track datasets. 
Given a handful of predictor variables that are common in either historical or simulated storm track ensembles such as pressure deficit at the storm's center, radius of maximal winds, storm center and direction, and distance to coast, the proposed stochastic model generates space-time fields of quantitative precipitation over the study domain. 
Statistically novel aspects include that the model is developed in Lagrangian coordinates with respect to the dynamic storm center that uses ideas from low-rank representations along with circular process models. 
The model is trained on a set of tropical cyclone data from an advanced weather forecasting model over the Gulf of Mexico and southern United States, and is validated  by cross-validation. 
Results show the model appropriately captures spatial asymmetry of cyclone precipitation patterns, total precipitation as well as the local distribution of precipitation at a set of case study locations along the coast. 

\end{abstract}

\section{Introduction} \label{sec:intro}

Tropical cyclones (TCs) are major drivers of coastal flooding. 
Due to the extreme nature of these events, quantifying flood risk is a difficult task that requires a combination of atmospheric, hydrologic and statistical modeling. 
Flood risk from TCs is primarily driven by two factors: storm surge due to intense winds over the open ocean, as well as substantial sustained precipitation over land. 
Storm surge causes flooding along the immediate coastline while extreme precipitation over land also causes flooding away from the coast, and the two hazards together conspire to cause extreme water levels throughout large areas that extend far from the TC center. 

Wind fields associated with tropical cyclones have been well-studied and are measured directly via aircraft- and satellite-based observing systems when TCs are located over the open ocean. 
However, much less is known about the forcing, distribution, and variation in tropical cyclone-induced precipitation over water and land \citep{elsberry2002}. 
Nevertheless, TC-induced precipitation is a key factor in defining the upper tail of flood distributions throughout the eastern United States \citep{villarini2010} and TC-induced precipitation is projected to increase in future storms due to changes in atmospheric moisture content and increasing storm intensities \citep{knutson2019}. 
Regardless of the impacts due to TC-induced precipitation, the lack of long observation records, the physical complexities associated with factors that control the precipitation, and large computational resources required to model such complex processes pose particularly difficult problems. 

There are four basic sources of data for TC precipitation: in situ observations, radar-derived estimates, satellite-derived estimates, and atmospheric model output. 
This study focuses on a statistical framework for generating complete space-time fields of TC-driven precipitation. 
Rain gauge data are not optimal for such a study, given their sparse spatial sampling and potential biases or measurement errors, especially in windy conditions such as during a TC \citep{villarini2008,pollock2018}. 
Radar-derived estimates such as those from Stage IV are lacking over open water where TCs form, and hourly estimates lack quality control \citep{nelson2016}. 
Satellite-derived estimates such as from the Global Precipitation Mission are gridded and available at high time frequencies, but have been found to underestimate heavy rainfall in coastal areas \citep{omranian2018}. 
This leaves atmospheric model output as a prime candidate for developing and training a statistical approach. 

Our approach to studying TC-driven precipitation is to exploit advanced atmospheric models such as the Weather Research and Forecasting (WRF) model developed at the National Center for Atmospheric Research \citep{skamarock2019}. 
Dynamic weather models such as WRF use physically-based equations to advance an atmospheric state forward through time, simulating the likely evolution of a number of weather-relevant features including temperature, winds, pressure, moisture and precipitation.  
These simulations prescribe and advance the ``state'' of the atmosphere as defined on a regular geospatial grid.

While global-scale weather simulations with these models are possible, they typically require enormous computational expense; for this reason scientists interested in spatially localized weather features such as TCs \citep{davis2008} typically use the WRF in a regional simulation mode, only simulating a small portion of the globe at a time.  
This still requires a fair amount of computational resources, but a reasonable-quality simulation of a single multi-day tropical cyclone event may be accomplished on a laptop within a few hours time \citep{hacker2017}.  
In addition, such simulations require initial and boundary conditions for the model that are consistent with the environment necessary to develop and maintain a tropical cyclone.  
These initial and boundary conditions (i.e., the temperature, wind, moisture and pressure at all grid points in the model) must maintain a dynamical balance to avoid instabilities in the model. 
In addition, it is difficult to constrain all the factors in the environment that lead to a TC developing to a particular strength or taking a particular path while retaining dynamical balance, making it hard to synthetically generate initial and boundary conditions in a systematic way.  
Rather, these are typically extracted from other, global scale forecasts or historical ``reanalyses'' of weather conditions which are often too coarse to well-resolve the tropical cyclone itself, requiring the WRF model to properly simulate the tropical cyclone structure.

Balanced global historical reanalyses of weather conditions suitable for constraining WRF simulations only go back to 1979 (e.g. ERA-Interim; \citealt{dee2011}) or, can be extended in some cases earlier in the 20th century (e.g., NOAA 20th Century Reanalysis v2, \citealt{compo2011}).  
This gives, at most, about 100 years of historical data with which to simulate tropical cyclone events.  
However, TC occurrence is a relatively rare event and 100 years of data record is not sufficient to provide adequate event samples to constrain the full distribution of possible tropical cyclone structures and impacts.  
Therefore, techniques are employed \citep{emanuel2006,hall2007,nakamura2015} to generate synthetic TCs to provide spatially and temporally resolved events to identify key distribution characteristics of TC occurrence.  
Generation of synthetic TCs are based on combined statistical methods and physically-relevant environmental conditions that govern formation, movement, and intensity of TCs over long observing periods in a much less computationally demanding manner than full physics, dynamical models.   
While synthetic TCs include track and intensity estimates, they do not account for TC-induced precipitation. 
Given the limitations induced by a relatively small set of historical cases in addition to the limiting capabilities associated with understanding, observing, and modeling TC-induced precipitation, the development of statistically-based ``emulators'' of TC precipitation becomes an attractive alternative.

TC-induced precipitation varies with storm size, intensity and track.  
While over open ocean, precipitation will vary with sea-surface temperatures and general atmospheric conditions such as vertical wind shear \citep{ueno2007}.  
When the TC approaches land and makes landfall, interactions with topography and increased friction will alter precipitation patterns and intensities.  
As noted above, the computational expense of using high-resolution models required to represent processes that govern TC-induced precipitation precludes the application of such models to routine estimates of TC-induced precipitation \citep{rogers2009}. 
Using large sets of satellite-based observations of TC-induced precipitation, \citet{lonfat2004} developed the Rainfall Climatology and Persistence (R-CLIPER) statistical model, of TC-induced precipitation patterns.  
This first-order model produced symmetric precipitation patterns about a TC center with no variation due to environmental interactions.  
\citet{lonfat2007} modified R-CLIPER to include asymmetries about the storm center caused by environmental interactions.  
While simple to implement, the R-CLIPER versions did not produce precipitation patters with sufficient detail to capture extreme variations due to environmental interactions, extreme storm intensity, or steep topography.  
A first model using simplified physical principles and detail in the atmosphere boundary layer was introduced by \citet{langousis2009}.  
This model explicitly modeled precipitation due to convergence of wind in the atmospheric boundary layer, which forces rising air and precipitation. 
While an improvement over R-CLIPER, the model of \citet{langousis2009} is only applicable over water as it does not account for topography or over-land friction. 
The precipitation model of \citet{zhu2013} improves upon the theoretical model of \citet{langousis2009} by building the tropical cyclone rainfall (TCR) model that incorporated topography, land-induced friction, and TC characteristics such as varying intensity and size. 
\citet{lu2018} compare TCR with full-physics, high-resolution simulations of TC-induced precipitation from WRF to determine that TCR produced detailed spatially- and temporally-varying precipitation patterns requiring only a fraction of the computation effort of WRF.  
While less expensive than WRF and still including important physical processes, the TCR does require representations of the TC and the environment in which it moves, which are not always available.

This paper focuses on developing a statistical framework for simulating TC precipitation fields as part of a larger effort to understand and quantify current and future coastal flood risk. 
While the stochastic generator can be trained on high-dimensional space-time fields of precipitation, a critical requirement is that the model must be able to generate precipitation given only a handful of standard univariate predictors that are commonly used in the TC risk community. 
Such predictors include general geophysical descriptors of a given TC behavior at any point in time, typically including location of the storm center, maximum wind speed (measure of intensity), direction and speed of motion, central pressure deficit with respect to the large-scale environment (a measure of strength) and radius of maximum winds (a measure of storm footprint). 
For each TC that occurs worldwide, such descriptors (and others) are archived in the International Best Track Archive for Climate Stewardship (IBTrACS, \citealt{knapp2010}). 
The IBTrACS database is constructed based on official TC analyses as defined by Regional Specialized Meteorological Centers (RSMCs) of the World Meteorological Organization (WMO).  
For example, North Atlantic hurricane entries into IBTrACS originate from the official warnings of the U.S. National Hurricane Center. 
While IBTrACS provides a source of actual worldwide TC activity over approximately the past 100 years (depending on ocean basin), synthetic TC data can be generated using a variety of methods \citep{emanuel2006,hall2007,nakamura2015} and are stored in formats that mimic the IBTrACS to enable consistency among data sets.

\begin{figure}[t]
  \centering
  \includegraphics[width=\linewidth]{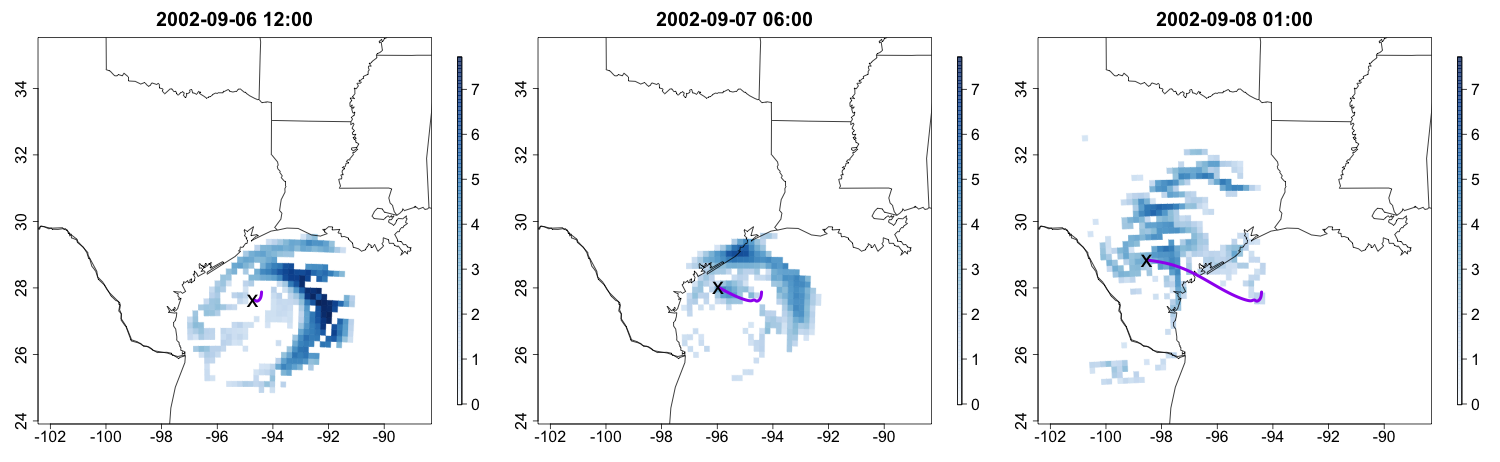}
  \includegraphics[width=\linewidth]{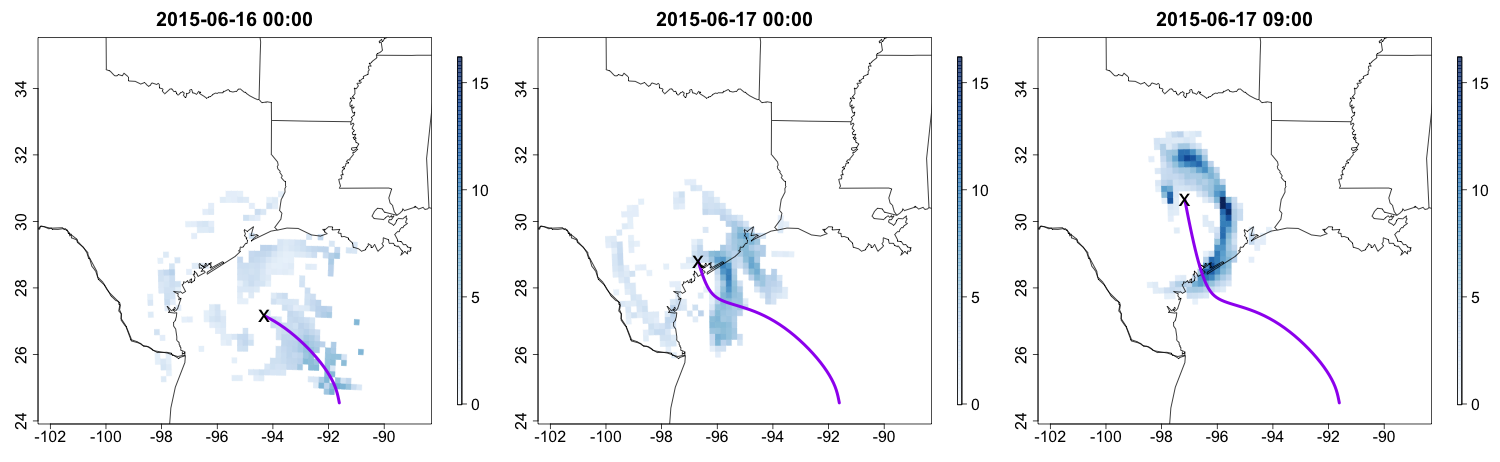}
  \caption{Each row contains example snapshots of hourly-integrated tropical cyclone precipitation; the purple line is the storm track to the current snapshot, while the ``x'' represents the storm center at the hour displayed.  The top row is hurricane Fay in September 2002, while the bottom row is hurricane Bill in June 2015.
  \label{fig:tc.example}}
\end{figure}

To understand and motivate the ensuing statistical approach, it is helpful to examine some partial realizations of tropical cyclones that directly impact the Texas coastline. 
Figure \ref{fig:tc.example} shows two such example tropical cyclone events from an atmospheric model detailed in the next section. 
Each row contains three snapshots of hourly-integrated precipitation along with the historical storm track, and estimated storm center. 
From simply a statistical point of view these images present some serious obstacles: precipitation exhibits a mixed discrete-continuous distribution whose spatial distribution changes in intensity and signature as functions of time. 
Moreover, interaction with land can lead to enhanced rainfall on the coast (e.g., the top central panel). 
Standard space-time statistical models are not well-suited for a non-Gaussian dynamical process such as those exhibited in Figure \ref{fig:tc.example}.

Our approach is to focus on the behavior of the storm, with respect to its estimated center. 
The precipitation patterns seen in Figure \ref{fig:tc.example} show a degree of spatial structure in a circular fashion with respect to the center of the storm. 
Modeling the process in polar coordinates presents some challenges, including the requirement of polar-indexed processes. 
The basic structure of the model is not too surprising for a reader familiar with the space-time statistics literature, including separation of mean and stochastic variation components. 
Key statistical innovations include a new polar-indexed process representation, as well as the incorporation of geophysical predictors in a statistical learning framework. 

\subsection{Data} \label{subsec:data}

To provide training data for this model, simulations of tropical cyclone events were performed with the Weather Research and Forecasting (WRF) model, which has been shown to produce realistic simulations of historical tropical cyclone events, including their precipitation characteristics (e.g., \citealt{davis2008,davis2010}. 
While there are some gridded observationally-based precipitation analyses available such as the National Center for Environmental Prediction (NCEP) Stage-IV precipitation analysis \citep{lin2005}, these analyses typically do not have many years of data (often only going back to the early 2000s) and suffer from data irregularities near and off of the coast as weather radar observation coverage diminishes.  
The WRF simulations allow the production of temporally and spatially consistent precipitation patterns for storms going back several decades.

To generate TC simulation training data, the WRF model is used to simulate seven historical TC events that impacted the Texas Gulf Coast, in particular hurricanes Fay (2002), Claudette (2003), Erin (2007), Humberto (2007), Edouard (2008), Ike (2008) and Bill (2015). 
The WRF model has a number of configuration options that can be tuned to change its performance, and such tuning was done here to optimize the WRF configuration to simulate these TC events. 
The exact WRF model configuration is described in the Appendix. 
Example snapshots of 1-hour accumulated precipitation output on the WRF grid is shown in Figure \ref{fig:tc.example}.

We additionally require a set of features that match information typically contained in storm track databases such as location of storm center, radius of maximal wind, central pressure deficit and storm direction.  
These quantities are calculated directly from the WRF output, and the derivations are detailed in the Appendix.

\section{Methods} \label{sec:methods}

Our interest focuses on hourly integrated precipitation amount at a spatial grid cell indexed by its center, $\bs$, at time point $t$.  
A standard approach would directly model such precipitation, denoted $P(\bs,t)=P(s_{lon},s_{lat},t)$.  
However, modeling in Euclidean coordinates is difficult due to the complex dynamic nature of tropical cyclones.  
In this section we describe the statistical model in generality, and in the ensuing section discuss the proposed estimation approach and choices specific to our case study.

\subsection{Statistical model} \label{subsec:overview}

We propose modeling $P$ in a Lagrangian coordinate system with respect to the space-time varying storm center. 
That is, we model $P(r,\theta,t)$ where $(r,\theta)$ are polar coordinates from the storm center at time $t$.  
Precipitation amounts are non-Gaussian and nonnegative; to this end we exploit a common integral transformation technique (sometimes called an anamorphosis or copula), 
\begin{align}  \label{eq:bias.correction}
  P(r,\theta,t) = 
  \begin{cases}
    G^{-1}(F(Y(r,\theta,t))), & Y(r,\theta,t) > 0 \\
    0, & Y(r,\theta,t) \leq 0
  \end{cases}
\end{align}
where $G^{-1}$ is a gamma quantile function, $F$ is a cumulative distribution function defined later and $Y(r,\theta,t)$ is a space-time Gaussian process. 
The model for $Y$ follows a familiar mean-plus-residual decomposition that is standard in the spatial modeling literature, but with details adapted to modeling space-time processes in polar coordinates:
\begin{align}  \label{eq:eof.plus.resid}
  Y(r,\theta,t) = \sum_{\ell=1}^L c_\ell(t) \varphi_\ell(r,\theta) + Z(r,\theta,t)
\end{align}
for time-varying stochastic coefficients $\{c_\ell(t)\}_{\ell=1}^L$, fixed basis functions $\{\varphi_\ell(r,\theta)\}_{\ell=1}^L$ and a space-time residual stochastic process $Z(r,\theta,t)$. 
The model (\ref{eq:eof.plus.resid}) assumes a stochastic mean function expressed as a weighted combination of spatial patterns.

Basis decompositions as in (\ref{eq:eof.plus.resid}) are very popular due to their computational benefits, as well as abilities to approximate processes from traditional spatial statistical models. 
They are also widely applicable and can be used to model nonstationary and heterogeneous processes  \citep{wikle2010,bandy2010,lindgren2011,cressie2008,nychka2015}. 
We first concern ourselves with specification of the mean function in (\ref{eq:eof.plus.resid}). 
Stochasticity is imposed in both the mean function coefficients and residual field, $Z(r,\theta,t)$. 
In particular, we decompose $\ell$th coefficient as 
\begin{align}  \label{eq:pc.model}
  c_\ell(t) = \mu_\ell(t) + W_\ell(t)
\end{align}
for a deterministic trend function $\mu_\ell(t)$ and time series process $W_\ell(t)$.

Exploratory data analyses do not suggest strong simple linear relationships between the predictors and the estimated coefficients, so we opt to model each $\mu_\ell(\cdot)$ as a random forest \citep{breiman2001} with predictors discussed in Section \ref{subsec:data}. 
The residual processes $W_\ell(t)$ are modeled as independent (across $\ell$), mean zero autoregressive Gaussian time series of order one.

\subsubsection*{Residual process}

It is worthwhile to distinguish the problem we are tackling in the context of process models for polar data. 
Much research has been devoted to directional statistics problems where the response is an angle or direction \citep{mardia2009}; a common model for univariate data is the Von Mises distribution.
\cite{jona-lasinio2012} extended such approaches to spatially-indexed data by introducing wrapped Gaussian processes. 
Our problem is distinct in that the response remains real-valued, but the indexing is no longer in Euclidean coordinates. 
The residual process at a given fixed band $r$ and time point $t$ must be periodic in $\theta$. 
We adopt a Gaussian process specification; due to the reference system these processes have recently been termed polar Gaussian process \citep{padonou2016}. 
Traditionally such a model would involve defining a positive definite function in polar coordinates; there are many routes to this end, \cite{gneiting2013} and \cite{rasmussen2006} discuss some approaches. 
Due to the moderate size of our dataset, rather than specifying a functional form of a covariance kernel, we opt for a constructive approach that ensures a nonnegative definite covariance kernel. 

To this end, we propose a basis representation that can be seen to be a synthesis of common low-rank ideas along with the polar Gaussian process idea. 
We model $Z(r,\theta,t)$ as 
\begin{align}  \label{eq:resid.model}
  Z(r,\theta,t) = \int k(r,\theta,r_0,\theta_0) U(r_0,\theta_0,t) 
  {\rm d}r_0 {\rm d}\theta_0
\end{align}
for a space-time correlated process $U(r,\theta,t)$.  
The model in (\ref{eq:resid.model}) can be seen to be an adapted case of the nonstationary mixture models of \cite{fuentes2001}, but different than the framework proposed by \cite{higdon1998} in that the kernel function does not vary over space, and $U$ is a correlated process.
The model for $U$ is a low-rank circular Gaussian process, 
\begin{align}  \label{eq:resid.model.2}
  U(r,\theta,t) = d_0(r,t) + \sum_{m=1}^M \left(d_{1,m}(r,t) \cos(m\theta) + 
  d_{2,m}(r,t) \sin(m\theta)\right)
\end{align}
where $\{d_i(\cdot,\cdot)\}$ are stochastic. 
The specification in (\ref{eq:resid.model.2}) can be interpreted as a harmonic decomposition for circular processes analogous to the spectral decomposition for stationary spatial processes \citep{stein1999}. 
The intercept accounts for any bias incurred from the fitted random forest mean functions, and is seen to provide more realistic residual field simulations. 
Each $d_0(r,t),\{d_{1,m}(r,t),d_{2,m}(r,t)\}_m$ is a radius-time process that is correlated in time according to an autoregressive process of order one, with a common autoregressive coefficient to avoid overfitting. 
For a fixed time point $t$, we model $\Cov(d_0(r_1,t),d_0(r_2,t))$ as a function of $r_1$ and $r_2$ nonparametrically (analogously for $d_{1,m}$ and $d_{2,m}$), described in the next section. 

\subsection{Estimation} \label{subsec:estimation}

Our estimation strategy is intimately tied to the way in which the model is developed. 
For example, is it difficult to envision a cohesive likelihood-based framework for simultaneously estimating empirical orthogonal functions, random forests defining the principal components and polar Gaussian process covariance parameters. 
Moreover, our objective is directly tied to simulating realistic precipitation fields rather than robustly estimating and interpreting statistical parameters. 
In the next section we show the stepwise approach works well in out-of-sample testing.

\subsubsection*{Data Description}

Our data $P_e(r,\theta,t)$ are available for $e=1,\ldots,n_e$ different TCs, or events, over hourly-spaced time points $t=1,\ldots,T_e$ the total number of which depends on the particular event. 
In this application $n_e = 7$ and the number of hours in a TC ranges between 72 and 108 hours; altogether there are $T=\sum_{e=1}^{n_e} T_e = 576$ observed realizations. 
The WRF data are originally available on a $59\times 59$ longitude-latitude grid, representing a 36 km spacing over the Gulf of Mexico region, extending into the southern United States (see e.g., Figure \ref{fig:case.study.map}).

\subsubsection*{Stochastic Mean Function}

We first concern ourselves with specifying and estimating the components of the stochastic mean function in (\ref{eq:eof.plus.resid}) and (\ref{eq:pc.model}). 
A common approach to modeling nonstationary processes in the atmospheric and climate sciences is to use an empirical orthogonal function representation \citep[EOFs;][]{vonstorch2002,wikle2010}, which is a low-rank representation in which the shape and form of the basis functions are specified by the data. 
Such an approach is equivalent to a principal components decomposition common in the statistical literature, and requires replicates of the process of interest on the same spatial grid. 
The native grid of our data in Euclidean space is common across all events and time points and directly lends itself to an EOF decomposition; however, the typical TC structure (e.g., the circular patterns of precipitation intensity about the center) varies widely across the domain depending on the location of the TC, which hampers an empirical method's ability to determine reasonable basis function structures. 
Moreover, the grid of our data in Euclidean space is common to all time points, modeling the process in a Lagrangian coordinate system results in irregularly-spaced and time-dependent coordinates in $(r,\theta)$ as the cyclone center moves across the domain. 

For the purpose of developing a mean function model, we artificially grid the Lagrangian-indexed data to a regular grid in polar coordinate space spanning $\theta\in[-\pi,\pi)$ and $r\in[0,1000]$ extending to 1000 km, resulting in a $100\times 100$ gridded product at each time point for each storm; denote this grid by $\{(r_{k_r},\theta_{k_\theta})\}_{k_r=1,k_{\theta}=1}^{100,100}$. 
The choice of gridding to $100$ points in each axial direction is admittedly arbitrary, but was seen to work as well as higher-order grids in exploratory simulations; moreover in radius this corresponds to a gridding of 10 km which is below the native resolution of the WRF data at 36 km.
For each event $e$ and time point $t$ we take the available $59\times 59$ data points $P_e(s_{lon},s_{lat},t)=P_e(r,\theta,t)$, fit a LatticeKrig model \citep{nychka2015} and krige to the regular grid in $(r,\theta)$. 
To account for differences of units, we rescale the radius axis to $[0,10]$. 
We use four levels of resolution and model settings that mimic an exponential covariance function, see \cite{nychka2015} for details on the LatticeKrig model. 
Because of the smoothing effect of kriging, the peaks of the original precipitation data will be slightly reduced in the polar-gridded dataset; in our experiments the median loss in maximal precipitation is 0.28 mm, or about $1\%$ under the max. 
This slight biasing is partially addressed in the anamorphosis transformation, discussed later.

Collecting together the gridded data in a $K\times T$ matrix $\bP$, where each column is a ``flattened'' polar image for all $T$ available time points, the EOFs  and principal components are obtained through a singular value decomposition (SVD) of $\bP$. 
In particular, if $\bP = \bU\bD\bV\T$ is the SVD of $\bP$ then the columns of $\bU$ represent ordered EOFs, and the columns of $(\bD\bV\T)\T$ the principal components.
These EOFs are spatial functions that define $\{\varphi_{\ell}(r,\theta)\}$.

We use the first 13 EOFs, which explain more than $75\%$ of variability. 
We also experimented with more EOFs, but found that the predictive ability of the principal component model, introduced next, suffered with higher order EOFs and likely resulted from overfitting. 
Instead, we opt for a parameterized residual process $Z(r,\theta,t)$.

The resulting principal components $\{c_\ell(t)\}_{\ell,t}$ are modeled as in (\ref{eq:pc.model}). 
The trend functions $\mu_\ell(t)$ are random forests \citep{breiman2001,hastie2009} with predictors of radius of maximal wind, central pressure deficit, storm center and direction (both vectors) and great circle distance of the storm center to coast. 
Some of these variables (e.g., storm center) are not directly available from the WRF output, but are derived to match the typical format of IBTrACS data; details can be found in the Appendix.
We use 500 trees and fit the forests using the \texttt{randomForest} package in \texttt{R} \citep{liaw2002}.
Estimated residuals are formed from the fitted trend functions $\hat{\mu}_\ell(t)$, $\hat{W}_\ell(t) = c_\ell(t) - \hat{\mu}_\ell(t)$. 
The covariance parameters of the first-order autoregressive model for $\hat{W}_\ell(\cdot)$ for a given event $\ell$ are estimated by maximum likelihood, yielding $\{\hat{\phi}_\ell,\hat{\sigma}_\ell^2\}_{\ell=1}^{n_e}$ where $\phi_\ell$ is the autoregressive coefficient and $\sigma_\ell^2$ is the innovation variance. 

\subsubsection*{Residual Process}

We turn to the residual process $Z(r,\theta,t)$, which follows a specification of (\ref{eq:resid.model}) and (\ref{eq:resid.model.2}). 
Based on the prior section, we have estimated trends $\hat{\mu}_\ell(t)$ from the fitted random forest. 
These yield estimated residuals $\hat{U}(r,\theta,t)=Y(r,\theta,t) - \sum_{\ell=1}^L \hat{\mu}_\ell(t) \varphi_\ell(r,\theta)$. 
Here it is useful to introduce additional notation:~call $\hat{U}(r_{k_r},\cdot,t)$ the estimated residuals for the $k_r$th radial band in the gridded polar coordinate system ($k_r=1,\ldots,100$). 
Using $\hat{U}(r_{k_r},\cdot,t)$ as data, coefficients  $d_0(r_{k_r},t),\{d_{1,m}(r_{k_r},t),d_{2,m}(r_{k_r},t)\}_m$ are estimated by least squares regression assuming a linear model of the form (\ref{eq:resid.model.2}), yielding $\hat{d}_0(r_{k_r},t),\{\hat{d}_{1,m}(r_{k_r},t),\hat{d}_{2,m}(r_{k_r},t)\}_m$. 
Taking $\hat{\bd}_{0}(t) = (\hat{d}_0(r_1,t),\ldots,\hat{d}_0(r_{100},t))\T$ we estimate the cross-covariance matrix for a fixed time point, $\Var\,\bd_0(t)$, empirically using $T^{-1}\sum_{t=1}^T \hat{\bd}_0(t)\hat{\bd}_0(t)\T$ (and analogously for $d_{1,m}$ and $d_{2,m}$). 
The autoregressive coefficient governing each of $d_0(r_k,\cdot)$ and  $\{d_{1,m}(r_{k_r},\cdot),d_{2,m}(r_{k_r},\cdot)\}_m$ are estimated independently by maximum likelihood assuming marginal Gaussianity, and are then averaged to provide a single autoregressive parameter estimate.

\subsubsection*{Transformation Function}

By this point all statistical parameters are estimated, except those governing the transformation functions $G$ and $F$ in (\ref{eq:bias.correction}). 
The parameters of the gamma function $G$ are estimated by maximum likelihood based on $\{P_e(r,\theta,t)\,|\,P_e(r,\theta,t) > 0\}$ assuming space-time independence. 
Other marginal transformations could be envisioned; if focus is on the extreme tail of the distribution then options that include generalized extreme value distribution tail behavior may be adopted \citep{naveau2016}; our initial focus is on the bulk of the precipitation distribution which is well modeled by the gamma transformation. 
Note the space-time varying nature of the statistical model implies that the distribution of precipitation at $(\bs_1,t)$ and $(\bs_2,t)$ are different, even with the common gamma transformation. 
Finally, $F$ is estimated as follows: assume an ensemble of $E$ realizations of space-time trajectories from the model for a given storm has been generated, $\{Y_e(\bs,t)\}_{e=1}^E$ on the original Euclidean grid and set of times for a study storm. 
We estimate $F$ as the empirical cumulative distribution function of the positive values $\{Y_e(\bs,t)\indicator_{[Y_e(\bs,t) > 0]}\}_{e=1}^E$, which, as $E\to\infty$, can be seen to approximate an equal weight mixture of local space-time densities; such an approach is an attempt to avoid overfitting of local probability density functions that can vary substantially over both space and time, and is seen to work well in the data study.

\subsubsection*{Simulation and Interpolation}

As modeling is done in regularly-spaced polar coordinates  $\{(r_{k_r},\theta_{k_\theta})\}$, initial simulations are quickly generated on the polar grid, yielding $\{U(r_{k_r},\theta_{k_\theta},t)\}$.  
For an arbitrary polar coordinate $(r,\theta)\not\in\{(r_{k_r},\theta_{k_\theta})\}$, the values of $Z(r_{k_r},\theta_{k_\theta},t)$ are bilinearly interpolated, i.e., $k(r,\theta,r_0,\theta_0)$ in (\ref{eq:resid.model}) represents a bilinear interpolation kernel. 
Generating $U(r,\theta,t)$ directly is not straightforward given the empirical estimation approach for the component coefficients $d_0,d_{1,m},d_{2,m}$. 
The EOFs are also bilinearly interpolated to the native grid for a given storm center, weighted by a realization of $\{c_\ell(t)\}$ and are added to $Z(r,\theta,t)$.
The final transformation (\ref{eq:bias.correction}) is applied at each grid cell separately. 

\section{Illustration} \label{sec:example}

In this section we explore the adequacy of the statistical model for capturing the distribution of precipitation patterns in the WRF data.  
First we explore some of the fitted model components based on the full available dataset, followed by detailed cross-validation study to assess the model's ability to capture out-of-sample behavior. 
In both cases we apply a tapering function to both the simulated and observed precipitation fields that is radially-symmetric and whose support depends on a particular time point's radius of maximal winds; the motivation for this is that our model is primarily designed to capture the local structure of tropical cyclones near the eye of the storm, whereas in WRF simulations other auxiliary precipitation events can be present due to other atmospheric conditions and forcings which our model is not designed to tackle. 
Future research may combine both tropical cyclone generators with other precipitation generators for convective or frontal events.

\subsection{Details of the model fit} \label{subsec:insample}

\begin{figure}[t]
  \centering
  \includegraphics[width=\linewidth]{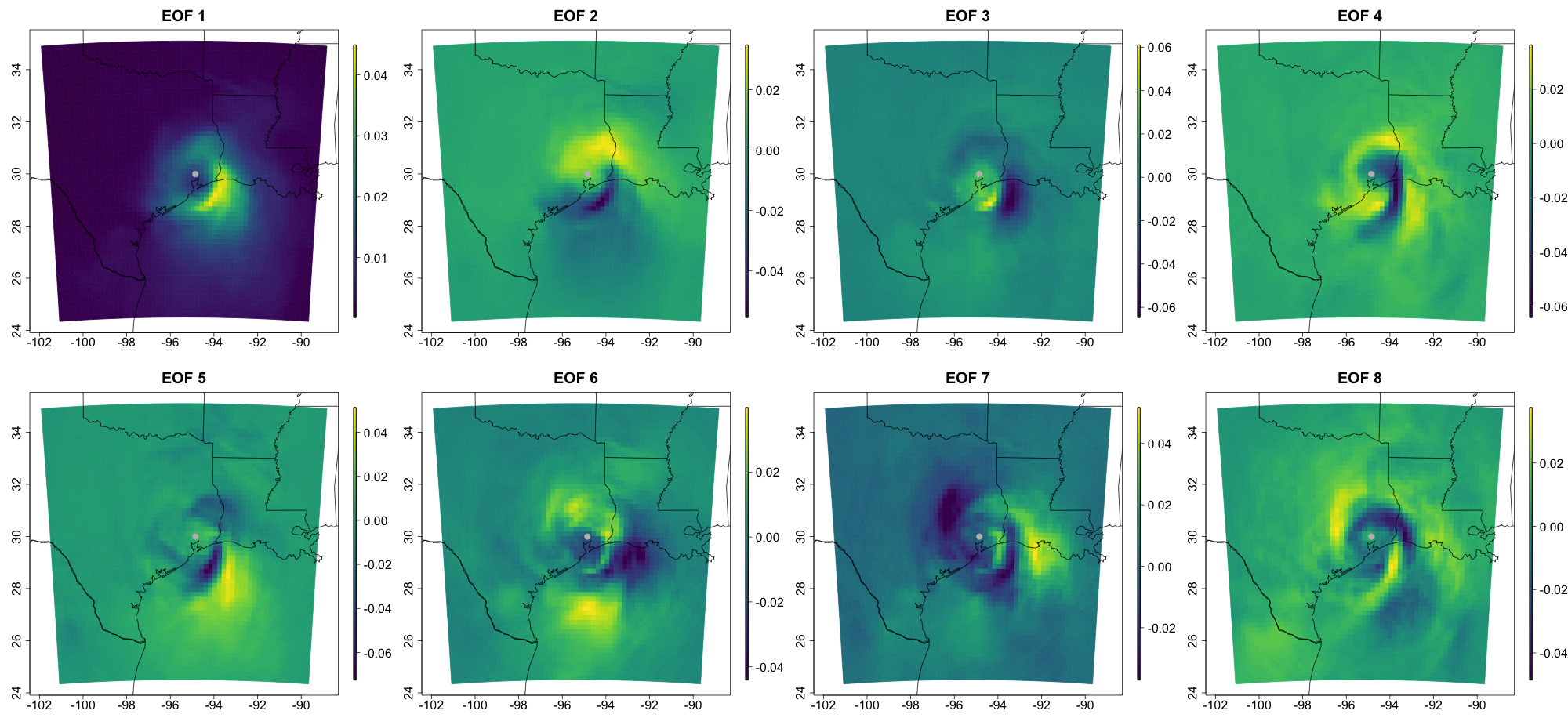}
  \caption{First eight empirical orthogonal functions the storm center indicated by a grey dot.
  \label{fig:eofs}}
\end{figure}

In this section we explore some details of the estimated model using the estimation procedure of the previous section applied to all seven tropical cyclones. 
The basis functions $\{\varphi_\ell(r,\theta)\}$ comprising the mean function in (\ref{eq:eof.plus.resid}) are empirical orthogonal functions, estimated as described in the previous section. 
Figure \ref{fig:eofs} shows the first eight estimated EOFs for a storm center located in approximately the center of the study domain. 
The first EOF shows a classical approximately-circular pattern around the storm center, but with a skewed feature to the southeast of the center. The second EOF can be seen to be a correction to the first that can reduce cyclone asymmetry and increase precipitation to the north of the center. The combination of these two EOFs with the coastline approximately determining the separation between positive and negative regions of these fields is anticipated, as there is often a notable enhancement of precipitation as the rain bands of the storm arrive over land. In addition, the concentric ``swirls'' seen in other EOFs relate to the banded nature of precipitation in tropical cyclones which is enhanced as they approach land.  The banding is enhanced as drier air from the continental interior gets wrapped into the center of the storm, suppressing precipitation in its path.

The coefficients of (\ref{eq:eof.plus.resid}) are modeled including a random forest component as in (\ref{eq:pc.model}). 
Although random forests consist of numerous easier-to-interpret trees, variable importance is one approach to assess the utility of features included in the predictor set over all trees simultaneously. 
Variable importance for a particular feature can be defined as the total (or average) reduction in variance explained at all nodes for all component trees over which that feature was split \citep{hastie2009}. 
Table \ref{tab:variable.importance} includes variable importance factors for the first seven principal components' random forest models. 
The first three principal components' random forest models highly favor the storm's central pressure deficit as the most important predictor, which is unsurprising given the statistical relationship between such deficit and the overall strength of the storm \cite{chavas2017}. 
As we move into higher frequency patterns, there are not apparent strong patterns highly favoring one particular feature over another. 
In the fourth principal component, the radius of maximum wind, V component of center (analogous to latitude) and distance to coast experience jumps in their variable importance factors, which indicates the importance of those variables for the EOF pattern seen in Figure \ref{fig:eofs}.

\begin{table}[tp]
\caption{Variable importance measured as total reduction in variance for the first seven principal components' random forest models; values are scaled by $1\times 10^{-5}$.
  \label{tab:variable.importance}}

\bigskip

\centering
\begin{tabular}{c|ccccccc}
  \hline
  \hline
  & $c_1$ & $c_2$ & $c_3$ & $c_4$ & $c_5$ & $c_6$ & $c_7$ \\
  \hline
  Radius of maximal winds    & 1.07 & 0.65 & 0.48 & 0.56 & 0.34 & 0.19 & 0.18 \\
  Pressure deficit at center & 9.97 & 4.61 & 1.06 & 0.70 & 0.89 & 0.64 & 0.61 \\
  U component of direction   & 1.57 & 1.06 & 0.78 & 0.47 & 0.41 & 0.29 & 0.24 \\
  V component of direction   & 3.82 & 0.84 & 0.98 & 0.69 & 0.45 & 0.51 & 0.51 \\
  U component of center      & 2.88 & 1.91 & 0.74 & 0.44 & 0.32 & 0.38 & 0.35 \\
  V component of center      & 1.86 & 1.10 & 0.61 & 0.96 & 0.40 & 0.32 & 0.29 \\
  Distance to coast          & 1.44 & 1.04 & 0.45 & 0.64 & 0.26 & 0.37 & 0.33 
\\ \hline
\end{tabular}
\end{table}

\subsection{Cross-validation} \label{subsec:cv}

In this section we consider the predictive ability of our approach. 
In particular, the model depends on only a handful of predictors: radius of maximal winds, central pressure deficit, storm center and direction of movement. 
Recall that these few predictors are those that are available in the widely-used IBTrACS data base or in other simulated products. 
With only these features, the model generates plausible precipitation patterns. 

To test the ability of the model to generalize beyond the few storms we have available, we consider a cross-validation experiment in which the estimation routine of Section \ref{subsec:estimation} is implemented on six of seven storms, the last of which is held out for comparison purposes. 
The estimation process is repeated for each possible hold-out storm, and in each case we generate an ensemble of 100 space-time realizations from the statistical model which we compare against the held-out WRF data.

\begin{figure}[t]
  \centering
  \includegraphics[width=\linewidth]{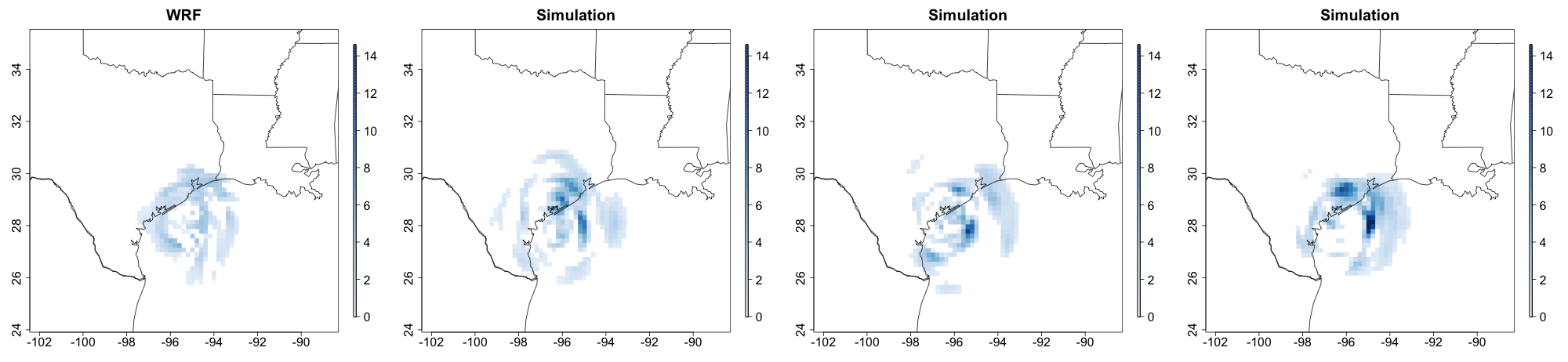}
  \includegraphics[width=\linewidth]{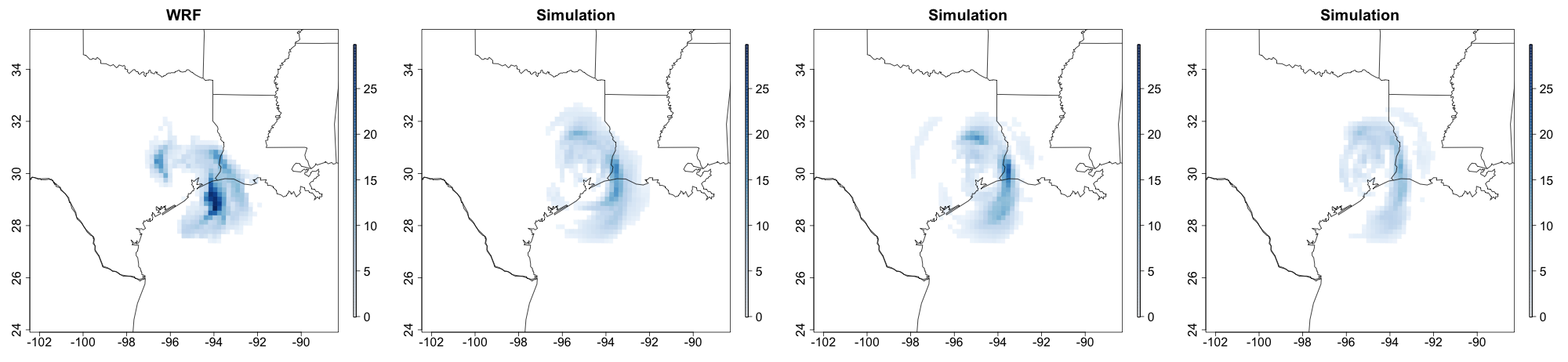}
  \includegraphics[width=\linewidth]{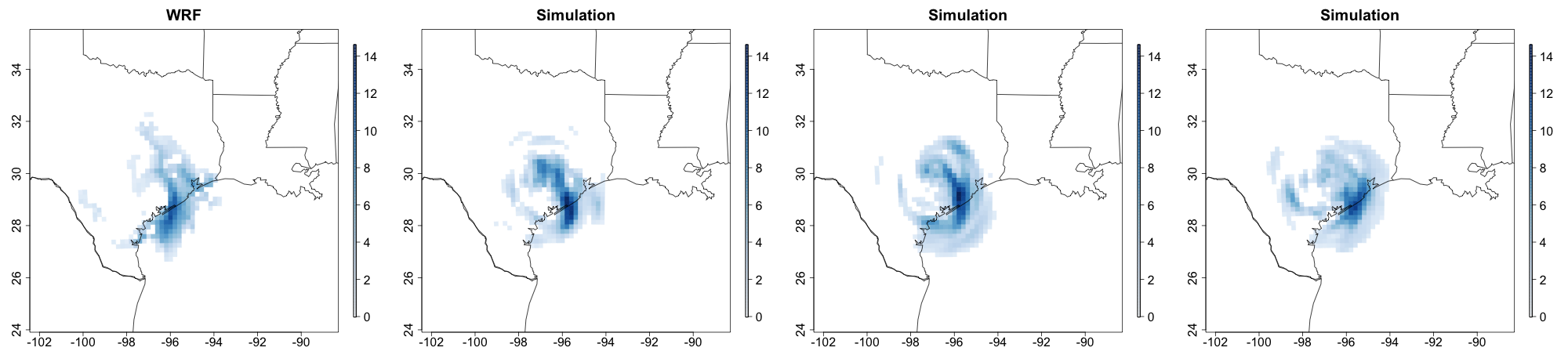}
  \caption{Cross-validation realizations for three snapshots. The three rows represent hurricanes Claudette on July 15, 2003 at 16:00, Ike on September 13, 2008 at 12:00 and Bill on June 17, 2015 at 04:00, respectively.  The first column is WRF output for the corresponding time, while the right three columns show cross-validation realizations from the statistical model. Units are mm.
  \label{fig:example.sims}}
\end{figure}

\subsubsection*{Space-time Structure}

To begin this section, Figure \ref{fig:example.sims} shows realizations from the  cross-validation experiment for a handful of instances. 
The three rows correspond to different TCs, the left column contains the held-out WRF simulation for a particular hour and the right three columns contain randomly chosen simulations from the simulated ensemble. 
We can see, for example in the second two instances, apparent asymmetry in the WRF fields that are replicated by the statistical model. 
In the first row, the first and second realization of the storm show heavier precipitation patterns, but that are more circular. 
The third realization shows stronger asymmetry with less precipitation. 

The asymmetrical patterns returned by the cross-validation realizations are consistent with observed patterns of tropical cyclone precipitation, particularly near landfall, where precipitation is enhanced to the right of the storm motion (here, to the east side of the storm).  
In addition, localized enhancements are seen in the portion of these rain bands immediately adjacent to the coast, particularly in the first and second row, which is also consistent with physical expectations \citep{rogers2009}.  
In the third row, this is a particularly difficult time to replicate, as the storm has moved well inland and is losing its structure as a tropical cyclone.  
Factors outside of the immediate environment of the tropical cyclone (including pre-existing atmospheric patterns over the continent) are more strongly influencing the structure of the decaying storm and increasing the variability in precipitation patterns away from the coast, and the statistical model struggles somewhat to capture this.  

\begin{figure}[t]
  \centering
  \includegraphics[width=\linewidth]{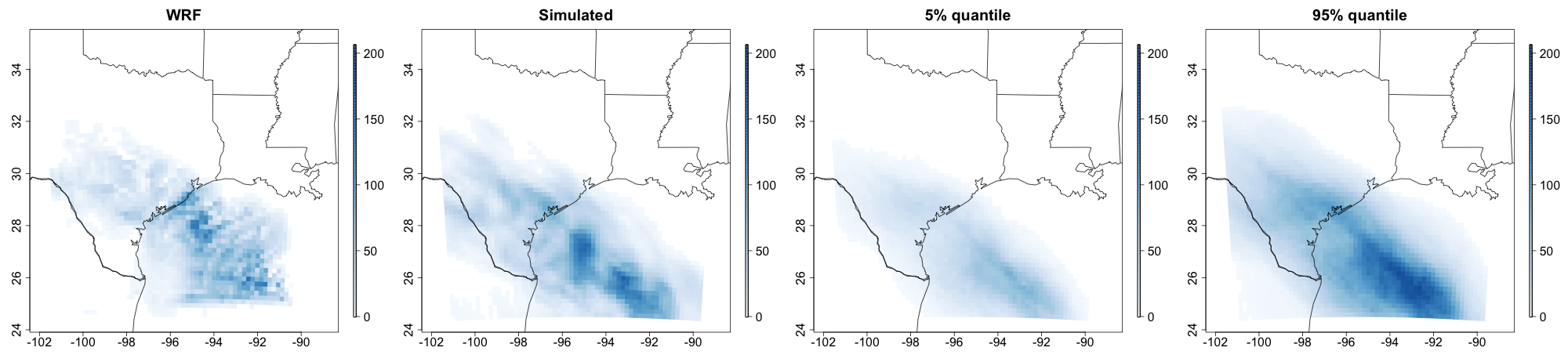}
  \caption{Integrated pointwise precipitation for the event beginning August 14, 2007 for WRF and a realization from the statistical model (left two panels), as well as pointwise $5\%$ and $95\%$ quantiles over an ensemble of 100 realizations of the statistical model. Units are mm.
  \label{fig:pixelwise.sum}}
\end{figure}

Although capturing the instantaneous spatial patterns of precipitation is important, it is also important to replicate temporally-aggregated behavior such as the total rainfall over the duration of the cyclone for flood risk modeling exercises. 
Figure \ref{fig:pixelwise.sum} shows pointwise-integrated precipitation over the study domain for one example event. 
The left panel represents integrated WRF and the second panel is one particular realization's integration; the next two panels show pointwise $5\%$ and $95\%$ quantiles of integrated precipitation over all realizations. 
First, the simulated totals appropriately capture the spatial distribution of total precipitation that cut in a diagonal pattern across the study domain. 
The statistical model exhibits some smoothing as compared to the actual WRF data, but capturing such high frequency behavior with our limited features is likely an unrealistic goal.

\begin{figure}[t]
  \centering
  \includegraphics[width=\linewidth]{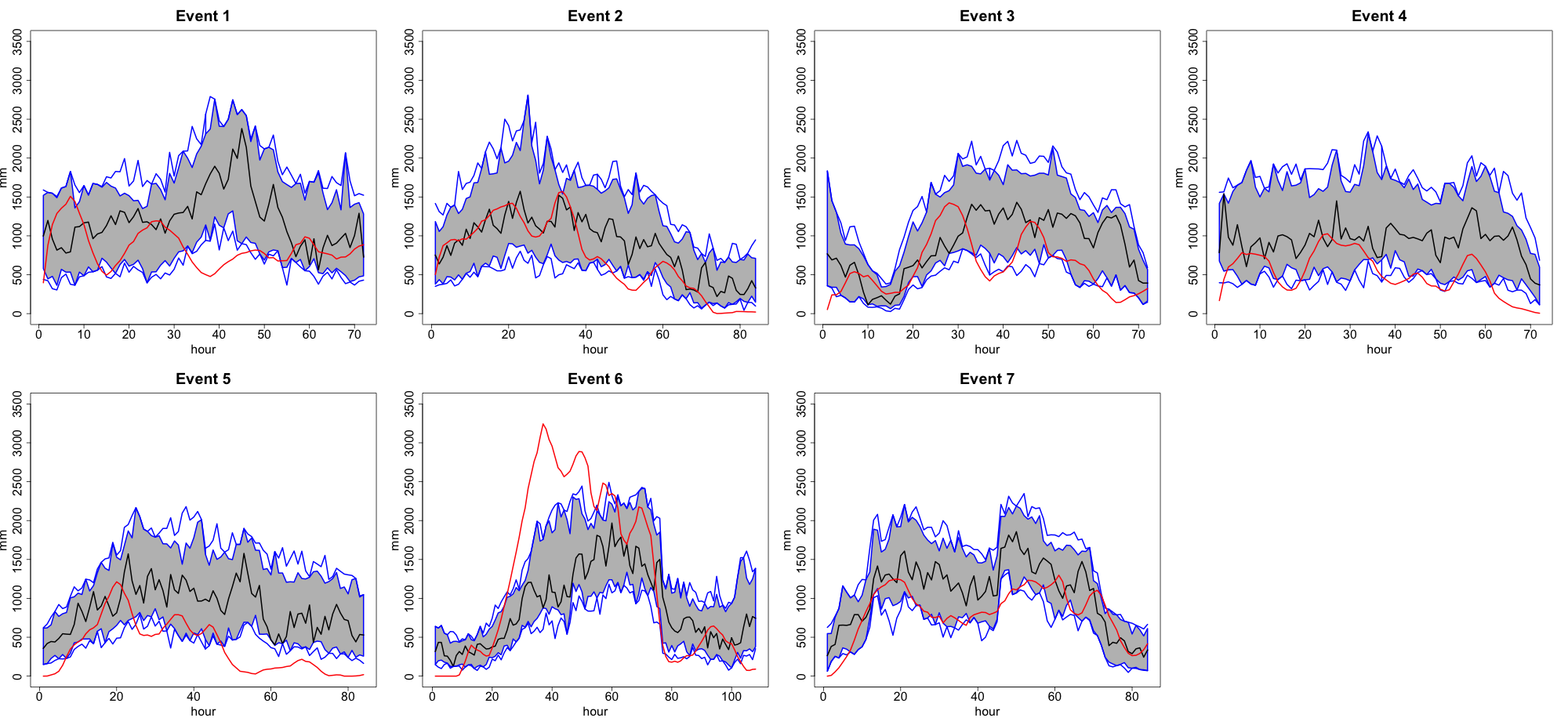}
  \caption{Functional boxplots of integrated precipitation over the study domain for each storm event based on an ensemble of 100 realizations of the statistical model.  The black line is the functional boxplot median while the red line is the integrated precipitation of WRF. Units are mm.
  \label{fig:domain.sum}}
\end{figure}

A third assessment of the model's ability to replicate space-time structure is in the temporal behavior of domain-wide integrated (i.e., total hourly) precipitation. 
Figure \ref{fig:domain.sum} shows functional boxplots \citep{sun2011} for integrated precipitation at each hour for each of the seven cross-validation storms. 
The red line indicates the held-out WRF statistic, while the grey band shows interquartile range of the WRF ensemble.
There is some additional high-frequency variation in the statistical model's integrated precipitation, where the WRF time series exhibits smoother variation of total precipitation than the statistical model. 
This could potentially be resolved by putting moderately-strong prior distributions on the autoregressive aspects of the model, encouraging greater temporal regularity, but is it not clear that the additional computational cost of such an implementation would be prohibitive. 
Generally the model realizations capture the verification time series within the $95\%$ functional boxplot confidence intervals, and also tend to follow the increasing and decreasing trends of precipitation. 
The sixth storm is an unusual or outlying case where substantially more precipitation fell over the domain (around hours 30-60) than in any of the other study storms; the statistical model captures the ramp-up period of increasing rainfall, but whose median does not adequately represent such intensity. 
This failure to capture the extreme integrated rainfall is due to the fact that the training storms (the other six) had more similar statistical characteristics with much less instantaneous precipitation falling on average. 

\subsubsection*{Local Verification and Case Studies}

\begin{figure}[t]
  \centering
  \includegraphics[width=0.5\linewidth]{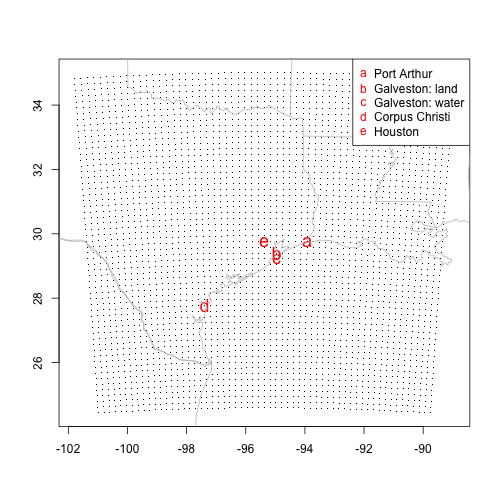}
  \caption{Locations of five case study locations on and near the Texas coastline; dots represent the computational WRF grid.
  \label{fig:case.study.map}}
\end{figure}

We now turn to local verification and case study examples of the statistical model. 
We consider five case study locations shown in Figure \ref{fig:case.study.map} on and near the Texas coastline which represents the primary area of interest for the original experimental design. 
The study locations are Port Arthur, Houston, Corpus Christi, TX and two pixels covering Galveston, TX: one pixel that is primarily over the land, the second is primarily over the ocean/coast adjacent to Galveston. 
The motivation for choosing the two Galveston locations is that precipitation patterns can change substantially with the introduction of land. 

\begin{figure}[t]
  \centering
  \includegraphics[width=\linewidth]{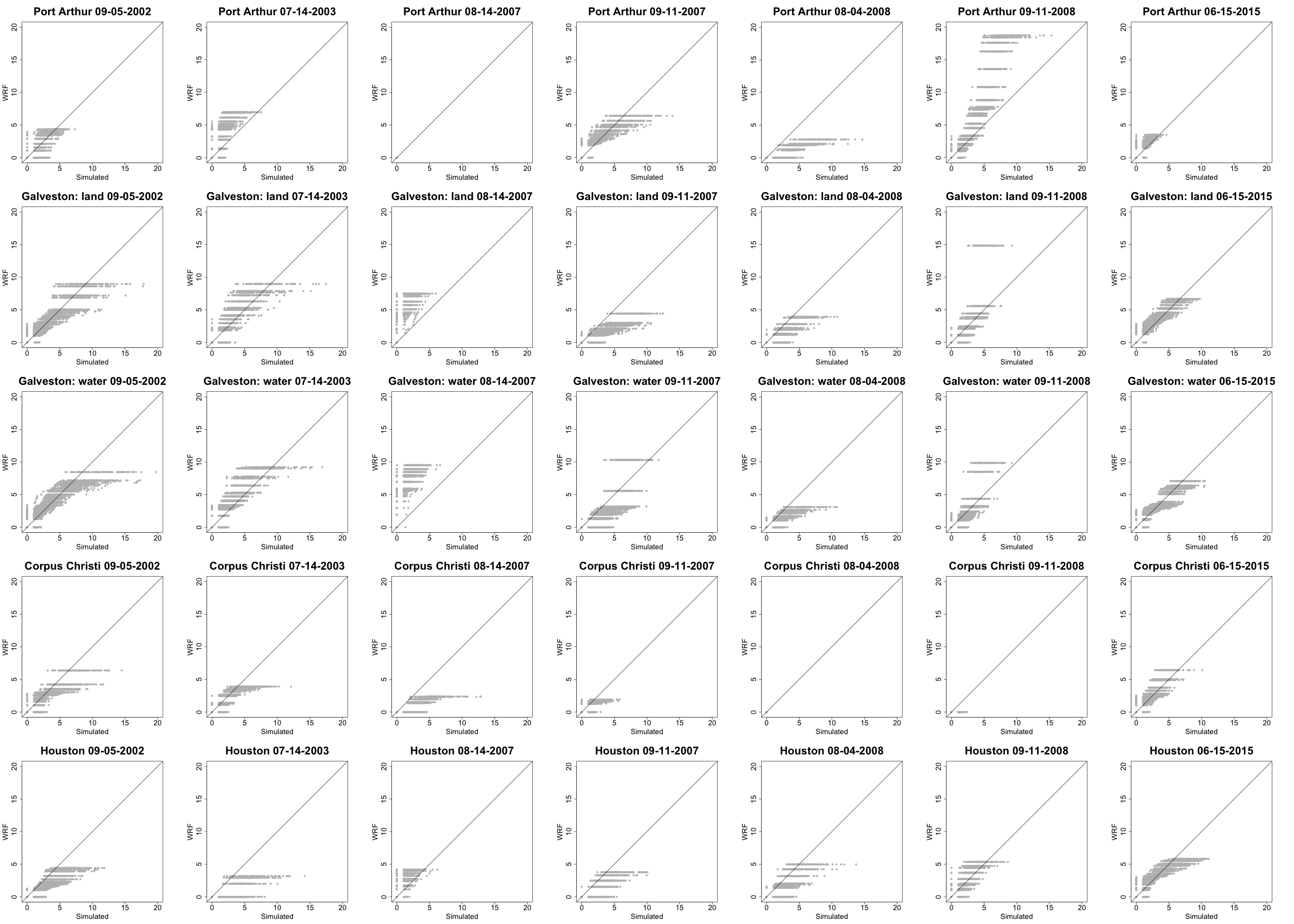}
  \caption{Q-Q plots of hourly precipitation at the five case study sites for all seven events.  Each row is a case study location and each column is a separate event. Units are mm.
  \label{fig:QQ}}
\end{figure}

We consider distributional aspects of hourly precipitation at each of the five case study locations in Figure \ref{fig:case.study.map} over all seven cross-validation events. 
Figure \ref{fig:QQ} shows Q-Q plots for each storm and location and each of the 100 realizations of the model. 
In particular, these Q-Q plots represent hourly precipitation data at the particular pixel of interest. 
These plots provide much information about the ability of the statistical model to appropriately localize precipitation. 
For example, in approximately six of the 35 Q-Q plots, the statistical model undersimulates precipitation at the extremes. 
Specifically, the sixth storm (column six) has high precipitation at Port Arthur and both Galveston pixels that are not adequately captured by the statistical model; as with Figure \ref{fig:domain.sum} this is likely due to the fact that this storm was statistically unusual in having substantially higher precipitation compared to the other events. 
Generally speaking the Q-Q plots typically contain the identity line in approximate $95\%$ confidence bands, even at more extreme precipitation amounts, although there are instances where the statistical model oversimulates precipitation (e.g., the first event at Houston, TX). 
Capturing such local distributions is a high demand on the model given the highly heterogeneous behavior of local precipitation, including its discrete-continuous nature. 
We next consider validating the occurrence of precipitation. 

\begin{figure}[t]
  \centering
  \includegraphics[width=0.4\linewidth]{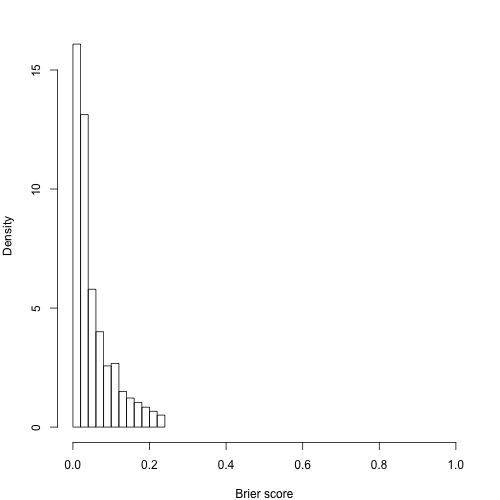}
  \caption{Histogram of Brier scores validating probability of precipitation over all hours and all TC events.
  \label{fig:brier}}
\end{figure}

The TC generator provides a separate probability distribution at each spatial location and hour; it is appropriate and useful to quantify the quality of these distributions with respect to the held-out WRF data using verification tools from probabilistic forecasting \citep{gneiting2007probabilistic}. 
In particular, we present two statistics, the first validating the model's ability to realistically provide probability of precipitation, and the second the full local simulation distribution.
A Brier score is a proper scoring rule used for predicting dichotomous events: for a set of forecast probabilities $f_i\in[0,1]$ and observations $o_i\in\{0,1\}$, the Brier score $i=1,\ldots,n$ is defined as 
\[
  \frac{1}{n}\sum_{i=1}^n (o_i - f_i)^2.
\]
Brier scores vary between zero and one, and are negatively oriented in that lower scores indicate better probability forecasts. 
We approximate the model's probability of precipitation by the empirical rate that the 100 ensemble members are nonzero at any given spatial location and hour. 
Figure \ref{fig:brier} shows a histogram of Brier scores for presence of positive precipitation over all hours from all events with almost all scores falling below $0.2$.
The average Brier score over all data is $0.05$, indicating high quality model probabilities of zero or positive precipitation that generally correctly identify locations of precipitation occurrence.

As a final verification statistic, we consider the full local predictive distributions of precipitation. 
In theme with validating the probability of precipitation, at each spatial location and time point there is a predictive distribution of precipitation sampled by the 100 realizations from the statistical model. 
Such predictive distributions can be validated using verification rank histograms \citep{gneiting2007probabilistic}. 
Rank histograms represent counts of the rank of the WRF precipitation (at a given pixel and time) within the 100 simulated values from the statistical model. 
If the model is well-calibrated, then these ranks should approximately follow a uniform distribution. 

\begin{figure}[t]
  \centering
  \includegraphics[width=\linewidth]{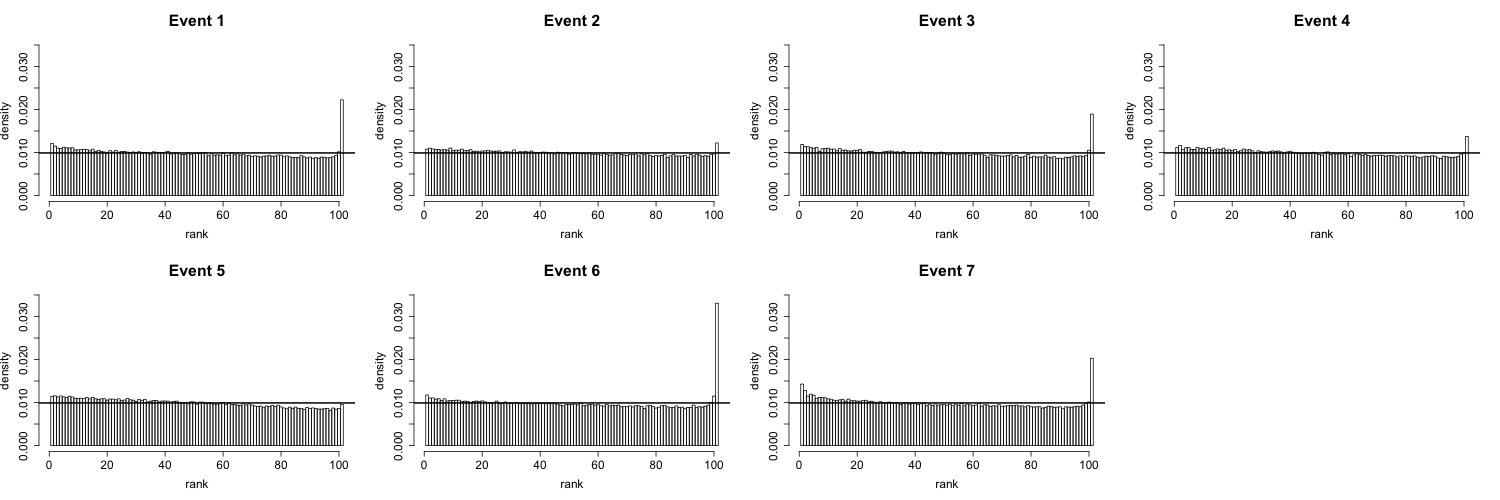}
  \caption{Verification rank histograms for each tropical cyclone event in the cross-validation experiment.
  \label{fig:PIT}}
\end{figure}

Figure \ref{fig:PIT} shows such rank histograms for each separate verification storm with a horizontal line indicating perfect calibration. 
Events with ties between the WRF data and simulated data are randomly disaggregated. 
Generally the histograms are flat and indicate high-quality calibration without much substantial under- or over-dispersion. 
Evidence shows in almost all cases of slight undersimulation at the most extreme events.  
That is, the presence of the peak at the highest rank indicates that there are an exaggerated number of cases where the WRF precipitation is higher than all 100 ensemble members from the model. 
These are not necessarily cases of extreme precipitation, although they are contributing factors. 
Future research may be to incorporate extreme value distributions into the space-time dynamical framework proposed in this paper.

\section{Discussion} \label{sec:discussion}

In this paper we introduce an approach to simulating quantitative precipitation fields for tropical cyclones given a handful of features that are typically available in historical or simulated tropical cyclone track databases. 
The method relies on modeling in Langrangian coordinates, using the estimated storm center as the coordinate system origin. 
Basis expansion models are used both within the mean and residual processes, exploiting links to standard kernel convolution approaches for nonstationary processes in the spatial statistical literature.

Estimation of the model is closely tied to model development, allowing for assumption checking and model building at each stage. 
A natural criticism of our approach is such a stepwise estimation procedure which is not expected to be statistically efficient. 
Taking a more principled likelihood-based approach would be desirable for a fully statistical solution, but in our experiments the current approach aligns well with the goal of this paper: to produce realistic space-time correlated fields of tropical cyclone precipitation. 
It is not clear that the added effort to more fully incorporate parametric uncertainty would have any substantive benefit, and would certainly come at the cost of significantly greater effort.

While our interest focuses on TC precipitation characteristics, coastal flooding is driven by both precipitation and storm surge. 
A future research direction is to fold in a wind field model, such as \cite{reich2007}, to estimate the joint hazard posed by these two aspects of TCs. 
Another relevant direction would be to experiment with the proposed approach in other geographical domains, as tropical cyclone climatological characteristics differ over distinct bodies of water.

\section*{Appendix} \label{apdx:WRF}

Here we describe the configuration of the WRF model and the extraction of predictors from the WRF model output.

\subsection*{WRF Model Configuration}

Version 4.0.1 of the Advanced Research WRF \citep[ARW;][]{skamarock2019} model is used to make these simulations.  
The WRF model has a number of settings that can be configured to change the exact formulation of the model; many of these settings are configured to simulate particular types of weather events.  
Several WRF model simulations with different configuration options were compared to hourly Automated Surface Observing System (ASOS) observations at major airports within the WRF domain, and the configuration that best matched the available in-situ observations was chosen for these simulations.  
The configuration options are outlined in the table below.

\begin{table}[h!]
  \caption{WRF model configuration, see text for details.}
  \bigskip
  \centering
  \begin{tabular}{c|cc}
  \hline\hline
  Configuration & Selection & Citation \\
  \hline
  Grid Spacing & 20 km & \\
  Initial and Boundary Conditions & ERA-Interim & \cite{dee2011} \\
  Microphysics & WSM5 & \cite{hong2004} \\
  Cumulus & Kain-Fritsch & \cite{kain2004} \\
  Planetary Boundary Layer & YSU & \cite{hong2006} \\
  Radiation & RRTMG & \cite{iacono2008}\\
  \hline
  \end{tabular}
\end{table}

\subsection*{Extraction of WRF Predictors}

The WRF output contains U and V direction wind speeds (in m/s) at multiple geopotential heights; we use speeds at 850 mb to derive both the storm center and radius of maximal winds. 
First, winds are spatially smoothed using a double exponential kernel which reduces aberrant artifacts in the derivation of predictors. 
We calculate the pointwise curl of U/V wind speeds using second-order discretized finite difference derivatives at each time step. 
At each hour, the storm center is identified as the grid cell with maximal curl. 
We restrict attention to storm centers that are within two degrees $L_\infty$ distance of the associated IBTrACS estimated storm centers to ensure that instantaneous aberrant storm center values are not used. 
Finally, the storm centers, which are registered to the WRF output grid, are then independently smoothed in both the U and V directions with a cubic smoothing spline.

Radius of maximal winds is calculated as the great circle distance to the point of maximal wind speed within 400 km of the estimated storm center at each hour. 
These radii are then smoothed over time with a cubic smoothing spline to reduce artifacts of the WRF gridding.
Storm direction is calculated at hour $t$ as the difference of the estimated storm center at time $t$ minus that at time $t-1$; such an estimate was found to be quantitatively similar to higher order approximations of direction. 
Central pressure deficit is quantified as 1013 minus pressure at the estimated storm center.


\end{document}